\documentclass[twocolumn,showpacs,superscriptaddress,groupedaddress,floatfix,preprintnumbers]{revtex4-1}
\usepackage{times}
\usepackage[pdftex]{graphicx} 
\usepackage{dcolumn}   
\usepackage{bm}        
\usepackage{amssymb}   
\usepackage{amsmath}   
\usepackage{natbib}
\usepackage{appendix}

\PassOptionsToPackage{caption=false}{subfig}
\usepackage{subfig}

\newcommand{\beq}{\begin{equation}}
\newcommand{\eeq}{\end{equation}}

\newcommand{\beqa}{\begin{eqnarray}}
\newcommand{\eeqa}{\end{eqnarray}}

\newcommand \nline {\nonumber \\}

\newcommand{\mbfr}{{\mathbf r}}

\newcommand \dxdy[2] {\frac{d #1}{d #2}}

\begin{document}

\title{Simulation of Early-stage Clustering in Ternary Metal Alloys Using the Phase Field Crystal Method}
\author{Vahid Fallah}
\affiliation{Mechanical and Mechatronics Engineering Department, University of Waterloo, 200 University Avenue West, Waterloo, Canada N2L-3G1}
\affiliation{Department of Materials Science and Engineering, McMaster University, 1280 Main Street West, Hamilton, Canada L8S-4L7}

\author{Nana Ofori-Opoku}
\affiliation{Department of Materials Science and Engineering, McMaster University, 1280 Main Street West, Hamilton, Canada L8S-4L7}

\author{Jonathan Stolle}
\affiliation{Department of Physics and Astronomy, McMaster University, 1280 Main Street West, Hamilton, Canada L8S-4L7}

\author{Nikolas Provatas}
\affiliation{Department of Physics, and Centre for the Physics of Materials McGill University, 3600 University Street, Montreal, Canada H3A-2T8}
\affiliation{Department of Materials Science and Engineering, McMaster University, 1280 Main Street West, Hamilton, Canada L8S-4L7}
\affiliation{Department of Physics and Astronomy, McMaster University, 1280 Main Street West, Hamilton, Canada L8S-4L7}

\author{Shahrzad Esmaeili}
\affiliation{Mechanical and Mechatronics Engineering Department,  University of Waterloo,200 University Avenue West, Waterloo, Canada N2L-3G1}

\begin{abstract}
Phase field crystal methodology is applied, for the first time, to study the effect of alloy composition on the clustering behavior of a quenched/aged supersaturated ternary Al alloy system. An analysis of the work of formation is built upon the methodology developed in Fallah {\it et al.} to describe the dislocation-mediated formation mechanisms of early clusters in binary alloys [Phys.~Rev.~B.,~DOI: 10.1103/PhysRevB.00.004100]. Consistent with the experiments, we demonstrate that the addition of Mg to an Al-1.1Cu alloy increases the nucleation rate of clusters in the quenched/aged state by increasing the effective driving force for nucleation, enhancing the dislocation stress relaxation and decreasing the surface energy associated with the Cu-rich Cu-Mg co-clusters. Furthermore, we show that it is thermodynamically favourable for small sub-critical clusters to have higher affinity for Mg than larger overcritical Cu-rich clusters, particularly depicting a two-stage clustering phenomenon.  
\end{abstract}

\maketitle

\section{Introduction}

The earliest stage of structural decomposition during quench/ageing of supersaturated solid solutions, referred to as \textit{solute clustering}, is a crucial step for establishment of their final microstructure. Solute clustering controls the mechanical properties of alloys through the dispersion pattern of small coherent lattice aggregates, namely clusters and/or GP zones. This phenomenon, also known as early-stage age hardening, is highly influenced by the chemical composition of the alloy, especially in multi-component systems. The physical mechanisms of clustering have been poorly understood by investigators exploring the effect of adding different elements to binary and ternary systems, as the atomistic behavior of the clustering phenomenon are often challenging to model or fully characterize at the atomic scale. 

Understanding the atomistic mechanisms of the clustering phenomenon in multi-component alloys is of crucial importance to efficiently design age-hardening processes for desired properties, and to accurately interpret the experimental observations and measurements. A systematic study of the clustering mechanisms precludes traditional atomistic methods, such as molecular dynamics (MD) and thermodynamic Monte Carlo (MC) simulations, which cannot operate on the diffusional time scales controlling clustering and related solid state transformations. Dynamical calculations with classical density functional theory (CDFT) also do not apply since they too operate on too small a time scale to be relevant to diffusion-controlled phase transformations processes ~\cite{jaatinen09}.  

Recently, a formalism coined the phase field crystal ({\it PFC}) methodology~\cite{elder04,elder07,wu10,greenwood10,greenwood11} has emerged that contains many of the salient fundamental principles of CDFT but which is suitably simplified so as to render calculations of microstructure kinetics with atomic-scale effects tractable on diffusive time scales. The atomic density field in the PFC formalism is coarse-grained in time \cite{jaatinen09} and does not have sharp peaks in solid phases allowing lower spatial resolutions. Numerous studies have reviewed and demonstrated the physics of the PFC methodology and its usefulness in describing a range of non-equilibrium microstructure phenomena, from solidification~\cite{tegze09,toth11} and grain boundary kinetics~\cite{wu12,greenwood12} to clustering~\cite{fallah12} and phase patterning due to atomic misfit strains~\cite{muralidharan10}. The most recent PFC formalism developed by Greenwood {\it et al.}~\cite{greenwood10,greenwood11} employs correlation kernels in the free energy that stabilize various crystal symmetries, and coexistence between them. More recently the approach was extended to binary alloys, represented by the dynamics of density and a concentration field~\cite{greenwood11-2}. 

In our most recent investigation~\cite{fallah12} with the binary PFC model of ref.~\cite{greenwood11-2}, we systematically elucidated a complete free energy path for early-stage clustering mediated by quenched-in dislocations in the bulk crystal, a mechanism inferred initially from previous experimental findings in binary alloys~\cite{babu12,ozawa70}. In particular, we showed that the energy barrier for formation of stable clusters can be lowered or even completely removed locally in the bulk matrix in the presence of an assembly of quenched-in dislocations. Here, we extend our energy analysis to ternary systems. The Al-Cu-Mg system is chosen since it has been vigorously studied for the evolution of clusters~\cite{marceau10-2,marceau10,somoza02,somoza00,nagai01,ringer97} and shown to exhibit enhanced clustering and age hardening by Mg alloying~\cite{marceau10-2,somoza02}. Moreover, the dominant effect of elasticity and, more particularly, the role of quenched-in dislocations has been observed during the early-stage decomposition of these alloys in quenched/aged state~\cite{nagai01}.   

In this paper we use a newly developed ternary phase field crystal model to study solute clustering phenomenon in ternary Al-Cu-Mg alloys. The details of a multi-component PFC model are presented in a separate paper. In this work, we focus on the effect of addition of a ternary species, i.e., Mg, on the clustering behavior of these alloys. The remainder of this paper is organized as follows. Section~\ref{TPFC} begins with an introduction of the ternary PFC model.  Section~~\ref{Eq} then demonstrates the model's equilibrium properties, explicitly focusing on the Al-Cu-Mg system. Section~\ref{Sim} then discusses new simulations showing the microstructural and compositional evolution of clusters. Section~\ref{En} details the analysis of system energetics in terms of cluster composition and work of formation during evolution. Where appropriate throughout Sections~\ref{Sim} and~\ref{En}, results are compared with experimental data in the literature.  

\section{Ternary PFC Model}
\label{TPFC}
This section reviews the main features of a simplified three-component free energy functional from which the dynamics of a phase field crystal density field and two impurity concentration fields are modelled. The starting point of our model is a multi-component analogue of the  classical density functional theory (CDFT) of freezing introduced by Ramakrishan and Yussouff~\cite{ramakrishan79}. Details of the derivation of the multi-component model are presented in a separate publication \cite{ofori12}. Only the details relevant to the ternary model used in this study are reproduced here. 

\subsection{Simplified ternary PFC Free Energy}
\label{simplified-PFC-multi-functional}

The free energy functional of a three-component system can be described by two contributions, ideal and excess energy, each as a function of three density fields (i.e., $\rho_{A}$, $\rho_{B}$ and $\rho_{C}$). From CDFT, we can write the following energy functional:
\begin{align}
\frac{\Delta {\mathcal F}}{k_B T \rho^o V } \equiv \int d\mbfr~{f}  = \int d\mbfr~\{\Delta F_{id}+\Delta F_{ex} \},
\label{truncated-DFT-energy}
\end{align} 
where $\Delta F_{id}$ are, respectively, the dimensionless the ideal energy, $\Delta F_{ex}$ and excess energy, where $k_B$ the Boltzmann constant,  $T$  the temperature, $\rho^o$ the average density (defined below) and $V$ the volume of the unit cell. The ideal free energy of the  mixture is given by 
\begin{align}
\Delta F_{id} &= \rho_{A} \ln \left(\frac{\rho_{A}}{\rho_{A}^{o}}\right)-\delta\rho_{A} + \rho_{B} \ln \left(\frac{\rho_{B}}{\rho_{B}^{o}}\right) - \delta\rho_{B}\nline&+\rho_{C} \ln \left(\frac{\rho_{C}}{\rho_{C}^{o}}\right) - \delta\rho_{C}, 
\label{idealenergy-log}
\end{align}  
where $\rho_i$ (with $i=A,B,C$) is the density of component $i$, $\delta \rho=\rho_i-\rho_i^o$ and ${\rho_{i}^{o}}$ the reference density of component $i$, taken to be that of liquid at solid-liquid coexistence. 

The excess energy term is described by a two-point correlations between atoms and introduces  elasticity, crystalline symmetry and gives rise to interactions between topological defects within solid phases. Considering particle interactions truncated to second order correlations in CDFT, this term can be written as
\begin{align}
\Delta F_{ex} &= -\frac{1}{2}\sum_i\sum_j \Delta F_{ij}\nonumber \\
&=-\frac{1}{2}\int d\mbfr^{\prime}\,\sum_i\sum_j\delta\rho_i\left(\mbfr\right)\,C_2^{ij}\left(\mbfr,\mbfr^\prime\right)\,
\delta\rho_j\left(\mbfr^\prime\right),
\label{two-particle-interaction}
\end{align}
where $C_2^{ij}$ denotes all possible combinations of the two particle correlations between components $i$ and $j$ with $i,j=A,B,C$. 

Following previous alloy PFC models~\cite{elder07,greenwood11}, we define a total mass density $\rho = \sum_{i} \rho_{i}$ and the total reference mass density as $\rho^o = \sum_{i} \rho_{i}^o$. Following Provatas and Majaniemi~\cite{provatas10} and Greenwood {\it et al.}~\cite{greenwood11}, the concentration of each component $i$ is defined as $c_{i} = \rho_{i}/\rho$ and the corresponding reference compositions are chosen as $c_{i}^o = \rho_{i}^o/\rho^o$. For convenience, we define a dimensionless mass density of the form $n = \rho/\rho^o-1$. From mass conservation, we have $\sum_i c_i \equiv 1$ (or $c_C=1-c_A-c_B$ and $c_C^o=1-c_A^o-c_B^o$). With these definitions, we re-write the free energy in terms of $n$ and the $\{c_i\}$. In doing so, some approximations are made for convenience. To avoid sharp density peaks, ideal free energy terms, expressed in $n$, are expanded to fourth order in the limit of small $n$. Also, we consider length scales where variations in concentration which are, to lowest order, slow compared to those of the density field, which varies on atomic scales.  In this limit, coarse-graining of the free energy or equations of motion, as in Refs.~\cite{majaniemi09,provatas10,elder10,huang10}, makes terms whose integrand is a function of  slow $c_i$ fields multiplying the fast $n$ field, vanish. Similarly, second order correlation terms containing combinations of the $c_i$ can be approximated by gradient terms in $c_i$.

With the above approximations at hand, the 3-component PFC energy functional for species $A$, $B$ and $C$ can be shown \cite{Ofori12} to reduce to
\begin{align}
\check{{\cal F}} &= \int d\mbfr~\Bigg\{\frac{n^2}{2} -\eta \frac{n^3}{6} + \chi \frac{n^4}{12} + \omega \, \Delta F_{\text {mix}} \, (n+1)
\nline&-\frac{1}{2}n\int d\mbfr^\prime C_{eff}(|\mbfr-\mbfr^\prime|)\,n^\prime + \frac{\alpha_{A}}{2}|\nabla c_{A}|^2 + \frac{\alpha_{B}}{2}|\nabla c_{B}|^2 \Bigg\},
\label{simplified-Energy}
\end{align}
where $\Delta F_{\text {mix}}$ denotes the ideal entropy of mixing,
\begin{align}
\Delta F_{\text {mix}} &= c_{A}\ln{\frac{c_{A}}{c_{A}^{o}}} 
+ c_{B}\ln{\frac{c_{B}}{c_{B}^o}}\nline &+(1-c_{A}-c_{B})\ln{\frac{(1-c_{A}-c_{B})}{1-c_{A}^o-c_{B}^o}}.
\label{entropy-mixing}
\end{align}
In Eq.~\ref{simplified-Energy}, $\eta$ and $\chi$ are parameters introduced to control the variation of the ideal free energy density away from the reference density $\rho^o$. A parameter $\omega$ is introduced to correct the entropy of mixing  away from the reference compositions $c_{A}^o$ and $c_{B}^o$. The gradient energy coefficients $\alpha_{A}$ and $\alpha_{B}$ set the scale and energy of compositional interfaces.  The parameters $\eta$, $\chi$, $\omega$, $\alpha_A$ and $\alpha_C$ have been shown to have contributions from from higher order correlative interactions~\cite{huang10}.   Here, we treat these parameters as free coefficients to match the free energy functional quantitatively to desired materials properties. 

The correlation function, $C_{eff}$, in Eq.~\ref{simplified-Energy} formally includes contributions from cross correlation functions of the form 
$C_{ij}$ in the excess energy. Extending the formalism of  Ref.~\cite{greenwood11} $C_{eff}$ to the case of ternary alloys, we define an effective correlation function in terms $C_2^{ii}$ according to
\begin{align}
C_{eff} &=  X_1 C_2^{AA} +  X_2 C_2^{BB} +  X_3 C_2^{CC},
\label{CorrEff}
\end{align}
where the coefficients $X_i$ are polynomial functions, which interpolate between two-body correlation kernels of the pure species, weighing each by the local compositions. The order of the coefficient $X_i$  vary depending on the number of components in the system and their order must be such as to smoothly interpolate from one correlation kernel to another. The coefficients satisfy $X_1+X_2+X_3 \equiv 1$ at all compositions. In this study they are defined by
\begin{align}
X_1 &= 1-3c_{B}^2+2c_{B}^3-3(1-c_{A}-c_{B})^2+2(1-c_{A}-c_{B})^3\nline &-4c_{A}c_{B}(1-c_{A}-c_{B})\nonumber \\[1.5ex] 
X_2 &= 1-3c_{A}^2+2c_{A}^3-3(1-c_{A}-c_{B})^2+2(1-c_{A}-c_{B})^3\nline &-4c_{A}c_{B}(1-c_{A}-c_{B})\nonumber\\[0.5ex]
X_3 &= 1-3c_{A}^2+2c_{A}^3-3c_{B}^2+2c_{B}^3-4c_{A}c_{B}(1-c_{A}-c_{B}).\nonumber \\[1.5ex] 
\label{interp-func}
\end{align}

The $\hat{C}^{ii}_2(\vec{k})$ are defined in Fourier space by peaks at $k_{j}$, which corresponds to the inverse of interplanar spacings of the main reflection from the $j^{\rm th}$ family of planes in the unit cell of the crystal structure favoured by component $i$. Each peak in reciprocal space is represented by the following Gaussian form of width $\alpha_j$, the height of which is modulated by a Debye-Waller-like prefactor, modulated by an effective temperature $\sigma$ and a transition temperature $\sigma_{Mj}$ ~\cite{greenwood11-2}, 
\beq
\hat{C}^{ii}_{2j}=e^{-\frac{\sigma^2}{\sigma^2_{Mj}}}e^{-\frac{(k-k_j)^2}{2\alpha^2_j}}. 
\label{CorrF}
\eeq
The $k=0$ mode of $\hat{C}^{ii}_{2j}$ have been omitted for simplicity but can be added through a constant. Its omission merely shifts the value of the average density and its effect can be modelled through the parameters of the model. 

\subsection{Model Dynamics}
\label{dynamics}
The dynamical equations of motion for each density field follow dissipative dynamics with stochastic noise in each field ~\cite{archer05}. When re-writing these equations in terms of a total density and two concentration fields, and neglecting the noise terms in each equation, we arrive at equations analogous to those used by Elder and co-workers in their alloy model \cite{elder07}, namely,
\begin{align}
\frac{\partial n}{\partial t} &= \nabla \cdot M_n\nabla \frac{\delta \check{{\cal F}}}{\delta\,n} \nonumber \\
&=\nabla\cdot M_n \nabla\Biggl\{n-\eta \frac{n^2}{2} + \chi \frac{n^3}{3} + \omega\Delta F_{\text {mix}}-C_{eff}\,n^\prime\Biggr\},
\label{dynamics-n}
\end{align}
\begin{align}
&\frac{\partial c_A}{\partial t} = \nabla \cdot M_{{c}_{A}}\nabla \frac{\delta \check{{\cal F}}}{\delta c_A}\nonumber \\
&=\nabla\cdot M_{{c}_{A}}\nabla\Biggl\{\omega(n+1)\frac{\partial \Delta F_{\text {mix}}}{\partial c_A}
-\frac{1}{2}n\frac{\partial C_{eff}}{\partial c_A}\,n^\prime - \alpha_A\nabla^2c_A \Biggr\},\nonumber \\
\label{dynamics-cA}
\end{align}
\begin{align}
&\frac{\partial c_B}{\partial t} = \nabla \cdot M_{{c}_{B}}\nabla \frac{\delta \check{{\cal F}}}{\delta c_B}\nonumber \\
&=\nabla\cdot M_{{c}_{B}}\nabla\Biggl\{\omega(n+1)\frac{\partial \Delta F_{\text {mix}}}{\partial c_B}
-\frac{1}{2}n\frac{\partial C_{eff}}{\partial c_B}\,n^\prime - \alpha_B\nabla^2c_B \Biggr\}.\nonumber \\
\label{dynamics-cB}
\end{align}

$M_n$, $M_{c_{A}}$ and $M_{c_{B}}$ are dimensionless mobility coefficients for density and compositions fields, which are, in principle, functions of the density, composition fields and temperature. In this study, at a given temperature, the mobility coefficients will be set to constants (i.e., equal to 1 for all mobilities).
  
\section{Equilibrium Properties}
\label{Eq}

We examined the equilibrium properties of a 2D Al-Cu-Mg system by constructing isothermal phase diagrams at two temperatures, i.e., $\sigma=0.155$ and $\sigma=0.04$, corresponding to solutionizing and quench/ageing temperatures, respectively. These temperatures are chosen to be well below the solid-liquid coexistence (solidus) temperature. The coexistence lines at each temperature are obtained by performing a common tangent plane construction on the free energy surfaces of solid phases expanded around the reference density ($\bar{n}=0$). The free energy surfaces of solid phases with square symmetry are calculated using a two-mode approximation of the density fields, given by, 
\begin{align}
n_i(\vec{r})=\sum_{j=1}^{N_i}A_j\sum_{l=1}^{N_j}e^{2\pi\mathbf{i}\vec{k}_{l,j}.\vec{r}/a_i},
\label{Density}
\end{align}
where the subscript $i$ denotes a particular solid phase with a lattice spacing $a_i$, and the index $j$ counts the number of modes $1 \cdots N_i$ in the $i$-phase. $A_j$ is the amplitude of mode $j$ and the index $l$ counts the $N_j$ reciprocal space peaks representing the mode $j$. $\vec{k}_{l,j}$ is then the reciprocal lattice vector corresponding to the index $l$ in family $j$, normalized to a lattice spacing of 1. 

The free energy surface for each phase can be calculated as a function of the two composition variables, $c_{Cu}$ and $c_{Mg}$, by substituting the above approximation of density field into Eq.~(\ref{simplified-Energy}) and integrating over the unit cell. This resulting free energy of the crystalline phase is then minimized for the amplitudes $A_j$. The minimization methodology is described in more detail in ref.~\cite{greenwood11-2,ofori12}.

\begin{figure}[htbp]
\resizebox{2.7in}{!}{\includegraphics{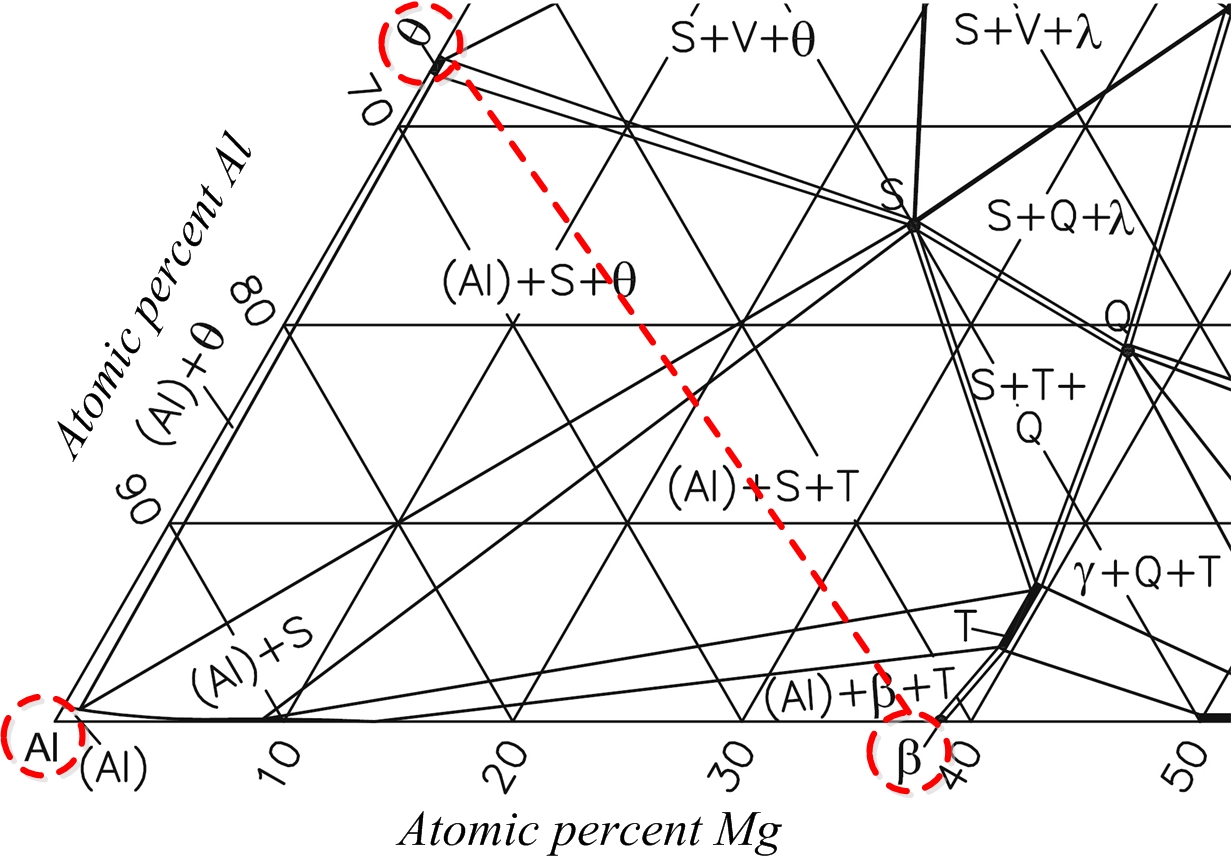}}
\caption{The Al-rich side of an isothermal cut (i.e., at $400^\circ C$) from the experimental phase diagram of Al-Cu-Mg system reprinted from Ref.~\cite{raghavan07}; Marked by the dashed circles are the Al-rich, $(Al)$, Cu-rich, ($\theta$), and Mg-rich, ($\beta$) phase concentration regions considered for reconstruction by the PFC model phase diagram, as shown in Fig.~\ref{fig:PhaseDiagram}. The dashed line represents the compositional boundary for Cu and Mg in our phase diagram calculations.}
\label{fig:PhaseDiagramExp}
\end{figure}

\begin{figure*}[htbp]
\resizebox{6.in}{!}{\includegraphics{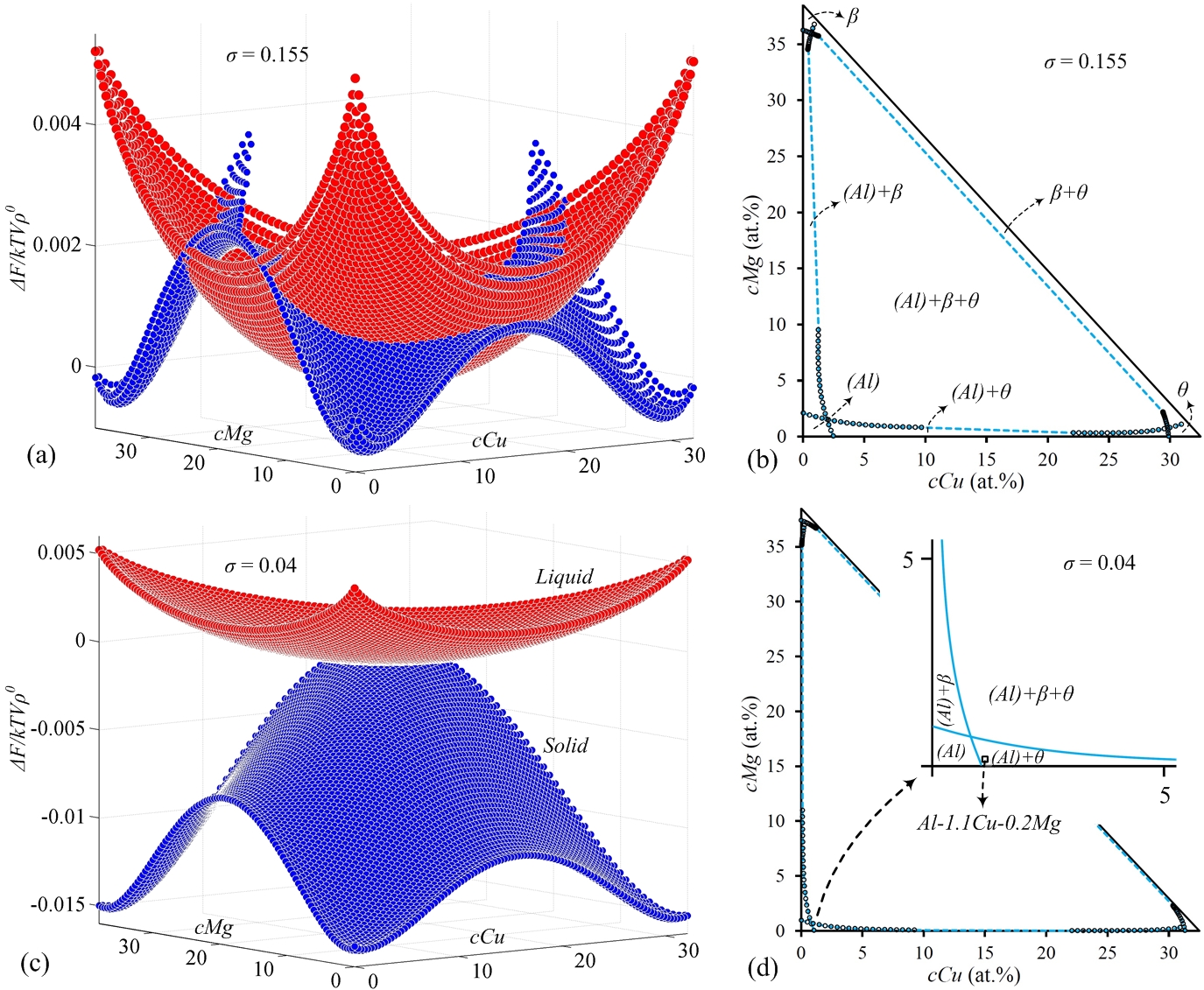}}
\caption{Solid and liquid energy landscapes of a square-square-square ($(Al)$-$\beta$-$\theta$) system at temperatures (a) $\sigma=0.155$ and (c) $\sigma=0.04$; Corresponding reconstructed phase diagrams at temperatures (b) $\sigma=0.155$ and (d) $\sigma=0.04$; The parameters for ideal free energy and entropy of mixing were $\eta=1.4$, $\chi=1$, $\omega=0.005$, $c_{Cu}^o=0.333$ and $c_{Mg}^o=0.333$. Widths of the correlations peaks are taken $\alpha_{11}=0.8$ and $\alpha_{10}=\sqrt{2}\alpha_{11}$ for all phases (the required ratio for isotropic elastic constants in a solid phase with square symmetry~\cite{greenwood11-2}). The peak positions are $k_{11(Al)}=2\pi$, $k_{10(Al)}=\sqrt{2}k_{11(Al)}$, $k_{11\theta}=(2.0822)\pi$, $k_{10\theta}=\sqrt{2}k_{11\theta}$, $k_{11\beta}=(1.8765)\pi$ and $k_{10\beta}=\sqrt{2}k_{11\beta}$. For simplicity, the effective transition temperatures $\sigma_{Mj}$ are set to 0.55 for all familes of planes in all phases; The concentrations $c_{Cu}$ and $c_{Mg}$ are rescaled considering the maximum Cu and Mg-content in the $\theta$-phase and $\beta$-phase according to the experimental phase diagram, i.e., $\approx 32.5$ and $\approx 38.5$ $at.\%$, respectively; The concentrations on the isothermal cuts are read in Cartesian coordinates.}
\label{fig:PhaseDiagram}
\end{figure*}

In the Al-rich corner of the experimental Al-Cu-Mg phase diagram, shown in Fig.~\ref{fig:PhaseDiagramExp}, there is a binary eutectic transition between the Al-rich $(Al)$-fcc phase and an intermediate phase $\theta$ (containing $\approx 32.5at.\%$ Cu) with a tetragonal crystal structure. The $(Al)$-$\theta$ system has a small solubility for Mg, reaching a maximum of $\approx 2at. \%$ Mg near the Al-rich side. Following the phase diagram in Fig.~\ref{fig:PhaseDiagramExp}, adding more Mg to the $(Al)$-$\theta$ system leads to the formation of a series of intermediate phases, such as $S$, $T$ and $\beta$. The latter is the cubic $\beta$-phase in the binary Al-Mg system with Mg-content of $\approx 38.5at. \%$. In our 2D model, we construct a phase diagramthat that maps onto the ternary eutectic system of $(Al)$-$\theta$-$\beta$, with all solid phases having square symmetry but differing in Cu and Mg-content. The lattice constant (and thus the reciprocal space peaks) of $\theta$ is interpolated between that of pure Al and Cu at $32.5at.\%$Cu. This calculation is performed also for the lattice constant of $\beta$ considering $38.5at.\%$Mg in the Al-Mg system. The free energy of the solid phase is generally calculated with a variable lattice constant weighted by concentrations $c_{Cu}$ and $c_{Mg}$ using the interpolation functions defined in Eqs.~\ref{interp-func}. The polynomial fitting parameters in Eq.~(\ref{simplified-Energy}) (namely $\eta$, $\chi$ and $\omega$) and the width of various peaks ($\alpha_j$) in the correlation kernel $\hat{C}^{ii}_{2j}$ are then selected so as to obtain approximately the same solubility limits for Cu and Mg in the solid phases as those in the experimental phase diagram. The parameters used are given in the caption of Fig.~\ref{fig:PhaseDiagram}.

Fig.~\ref{fig:PhaseDiagram} shows the free energy landscapes of solid and liquid along with the corresponding phase diagrams constructed for solutionizing and ageing temperature parameters of the model, i.e., $\sigma=0.155$ and $\sigma=0.04$, respectively. To construct the isothermal phase diagrams, the coexistence (solidus) lines for $(Al)-\theta$, $(Al)-\beta$, $\beta-\theta$ and $(Al)-\beta-\theta$ were obtained by requiring that the chemical potential and grand potential are equal for each species in the chosen phases. For example, the following set of equations were solved to find the $(Al)-\theta$ coexistence line: 

\begin{align}
&\mu_{c_{Cu}}^{(Al)}=\mu_{c_{Cu}}^{\theta}\nline&\mu_{c_{Mg}}^{(Al)}=\mu_{c_{Mg}}^{\theta}
\nline& f^{(Al)}-\mu_{c_{Cu}}^{(Al)}c_{Cu}^{(Al)}-\mu_{c_{Mg}}^{(Al)}c_{Mg}^{(Al)}=f^{\theta}-\mu_{c_{Cu}}^{\theta}c_{Cu}^{\theta}-\mu_{c_{Mg}}^{\theta}c_{Mg}^{\theta},
\label{coex}
\end{align}  
where $\mu_{c_{Cu}}=\partial f/\partial {(c_{Cu})}$ and $\mu_{c_{Mg}}=\partial f/\partial {(c_{Mg})}$ are the chemical potentials of the concentrations $c_{Cu}$ and $c_{Mg}$, respectively. Figure~\ref{fig:PhaseDiagram}(d) shows good agreement, in terms of the solubility of Mg, of the single-phase $(Al)$ and multi-phase $(Al)-\theta$ structures, in the dilute-Mg part of the experimental phase diagram of Fig.~\ref{fig:PhaseDiagramExp} and that of the constructed one at $\sigma=0.04$. 

\section{Clustering Simulations}
\label{Sim}

This section presents simulation results of the clustering phenomenon in Al-Cu-Mg system in the form of microstructural and compositional evolution of clusters, and results are compared with relevant experimental evidence gathered from the literature. 

Using the calculated equilibrium properties discussed in the previous section, simulations of clustering were performed on a 2D rectangular mesh with grid spacing $dx=0.125$ and time step $dt=10$. The size of the grid was $4096 \times 4096$ gird spacings (equivalent to $512 \times 512$ atoms). Each atomic spacing was resolved by 8 mesh spacings considering lattice parameter of 1 for a 2D square structure. To solve the dynamical equations, a semi-implicit algorithm was used in Fourier space for higher efficiency. The initial conditions were chosen to include the quenched-in dislocations in the bulk crystal, which are proposed to play a dominant role during the early stage clustering in quenched/aged Al-Cu and Al-Cu-Mg alloys~\cite{babu12,ozawa70,ringer96,nagai01,marceau10}. Following Fallah {\it et al.}~\cite{fallah12}, initial conditions employed a a crystal lattice of uniform composition, which is distorted at the quench temperature by introducing a uniform distribution of dislocations. As with the previous study~\cite{fallah12}, the character of dislocations is not of our concern in this 2D study, and they are simply defined with edge dislocation characteristics, with the dislocation line perpendicular to the surface. While, in this study, we investigate the dislocation-mediated mechanisms of clustering, we expect that the proposed mechanisms will also hold for vacancy-assisted clustering.    

\begin{figure*}[htbp]
\resizebox{7in}{!}{\includegraphics{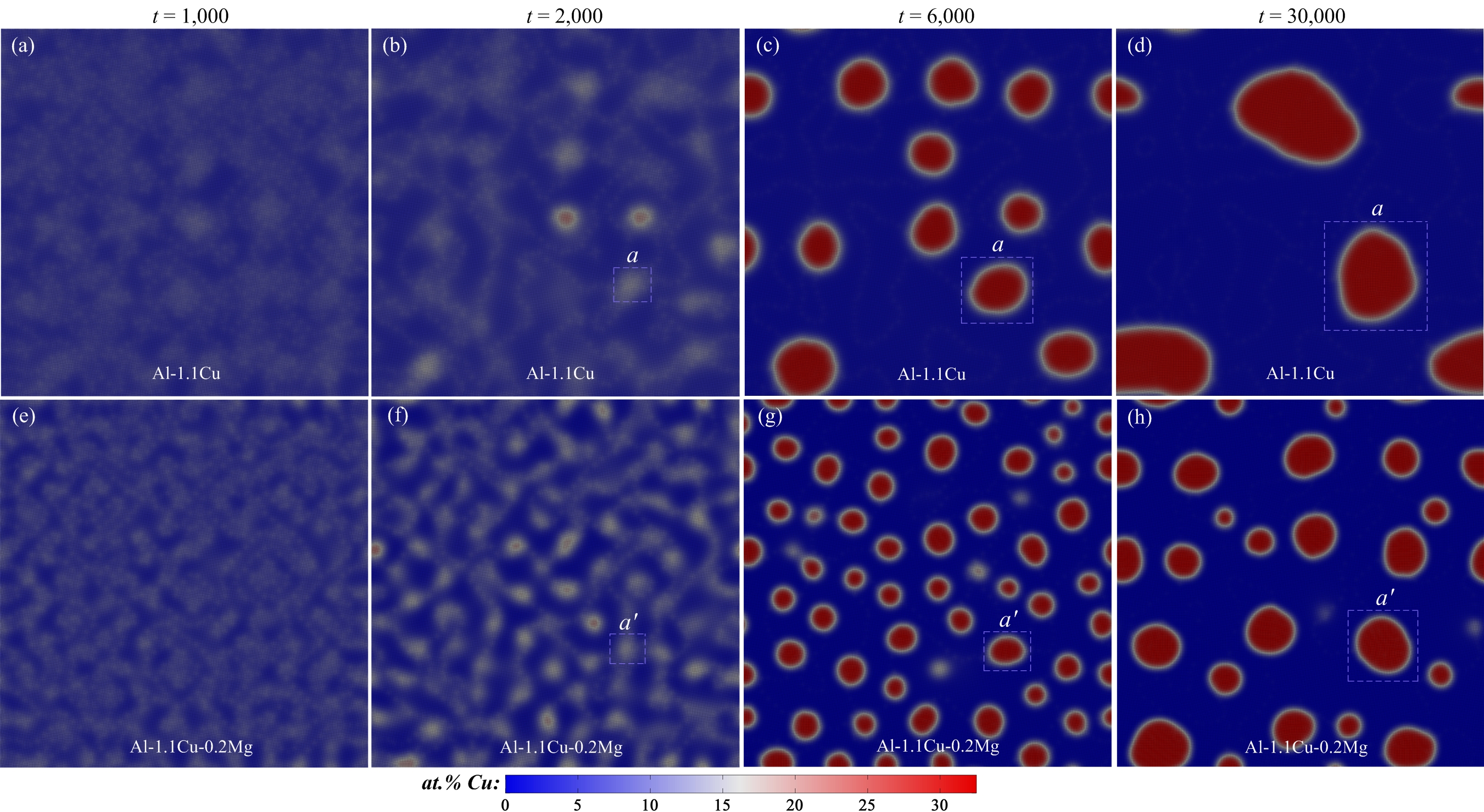}}
\caption{Time evolution of clusters in solutionized/quenched (a)-(d) Al-1.1Cu and (e)-(h) Al-1.1Cu-0.2Mg alloys at model temperature $\sigma=0.04$; Typical stable clusters ``$a$'' and ``$a'$'' are labelled on the microstructures of the Al-1.1Cu and Al-1.1Cu-0.2Mg alloys, respectively; Each image represents a simulation domain with $512 \times 512$ atoms.}
\label{fig:Clustering}
\end{figure*}

\begin{figure*}[htbp]
\resizebox{5.2in}{!}{\includegraphics{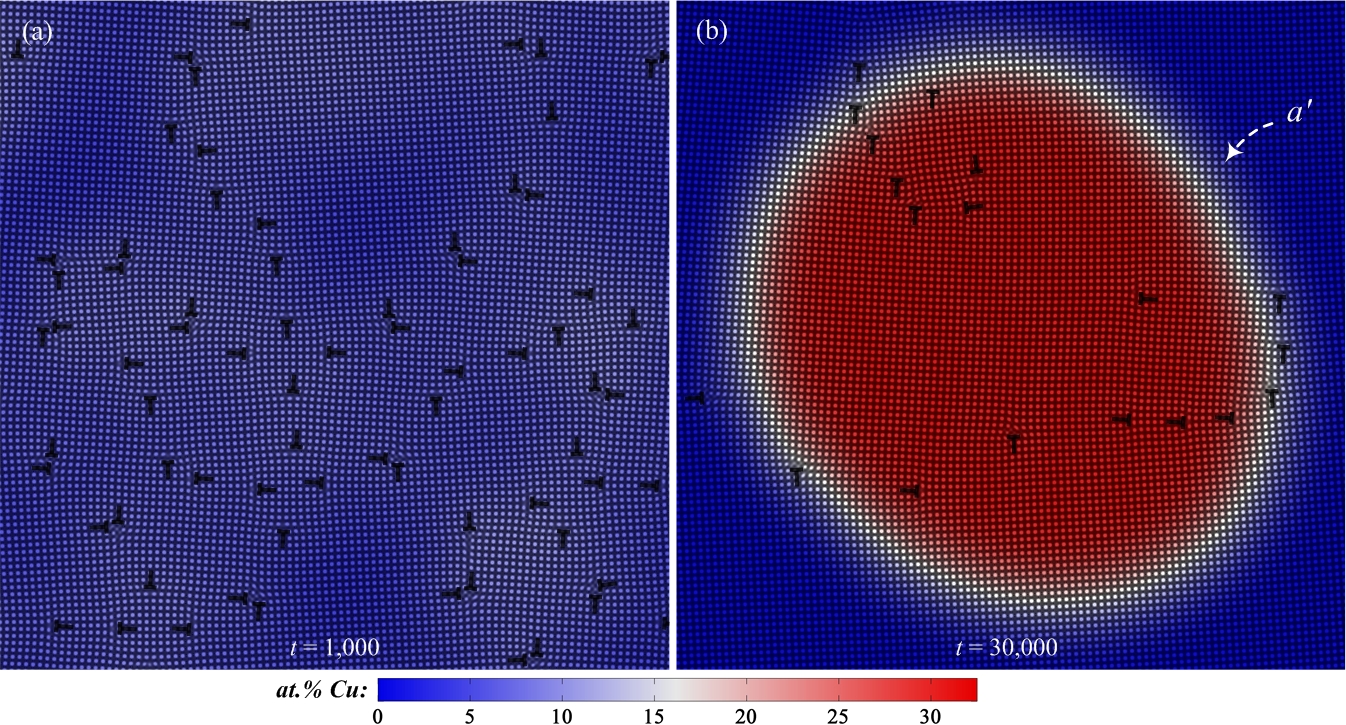}}
\caption{Zoomed-in images of the area within the box labelled Cluster ``$a'$'' (in Fig.~\ref{fig:Clustering}(h)) in the Al-1.1Cu-0.2Mg alloy at (a) $t=1,000$ and (b) at $t=30,000$ time steps; Black dots indicate atomic positions and T-like symbols label dislocations.}
\label{fig:ClusteringStructure}
\end{figure*}

Fig.~\ref{fig:Clustering}(a)-(d) and (e)-(h) show the PFC simulation results for quench/aging of Al-1.1Cu and Al-1.1Cu-0.2Mg, respectively, from the solutionizing temperature of $\sigma=0.155$ to $\sigma=0.04$. Labelled on these images are the typical stable clusters ``$a$'' and ``$a'$'' in Al-1.1Cu and Al-1.1Cu-0.2Mg alloys, respectively, that survived the growth competition among the other clusters. As can be inferred from Fig.~\ref{fig:Clustering}, the number of observed clusters within the unit area of simulation, at each time step, is much larger for Al-1.1Cu-0.2Mg alloy compared to Al-1.1Cu alloy. The zoomed-in images of the area within the box labelled Cluster ``$a'$'' in Fig.~\ref{fig:Clustering}(h) is shown in Fig.~\ref{fig:ClusteringStructure}(a) and (b) at time steps $t=1,000$ and $t=30,000$, respectively. Fig.~\ref{fig:ClusteringStructure}(a) shows how initially the Cu atoms segregate into the areas around the dislocations. Over the simulation time, as illustrated in Fig.~\ref{fig:ClusteringStructure}(b), the system undergoes a process of rearrangement and/or annihilation of dislocations within the matrix, leading to the formation of Cluster ``$a'$''.      

The selected alloy compositions for this study (i.e., Al-1.1Cu and Al-1.1Cu-0.2Mg) are within the single-phase region of $(Al)$ at the solutionizing temperature, while they are located within the two-phase region of $(Al)$-$\theta$ at the quench/ageing temperature (i.e., $\sigma=0.155$ and $\sigma=0.04$, referring to the phase diagrams in Fig.\ref{fig:PhaseDiagram}(b) and (d), respectively). With these concentrations, during clustering simulations, we expect to observe Cu-rich clusters evolving with the composition of the equilibrium $\theta$-phase (as can be seen on the final microstructures shown on Fig.~\ref{fig:Clustering}(d) and (h)). Also, having set the square symmetry for all species, the lattice parameter will be the only structural factor expected to change as a cluster evolves in the matrix during the 2D clustering simulations. Intermediate non-equilibrium phases in Al-Cu-Mg system, which are not predicted by the experimental phase diagram, were not  considered in this study.     

During the simulation of ageing process in both alloys, first small clusters form with a slightly higher Cu and Mg-content than that of the matrix. As ageing progresses, some of these clusters shrink in size and become depleted in Cu and Mg, but a few become stabilized (e.g. all the clusters shown in Fig.~\ref{fig:Clustering}(d) and (h), and typically those labelled ``$a$'' and ``$a'$'' for Al-1.1Cu and Al-1.1Cu-0.2Mg alloys, respectively). In contrast, ageing at a temperature within the single-phase $(Al)$ region for both alloys, i.e., $\sigma=0.145$, results in complete elimination of distortion, as expected. 

\subsubsection{Effect of Mg on the evolution of microstructure }
\begin{figure}[htbp]
\resizebox{3.3in}{!}{\includegraphics{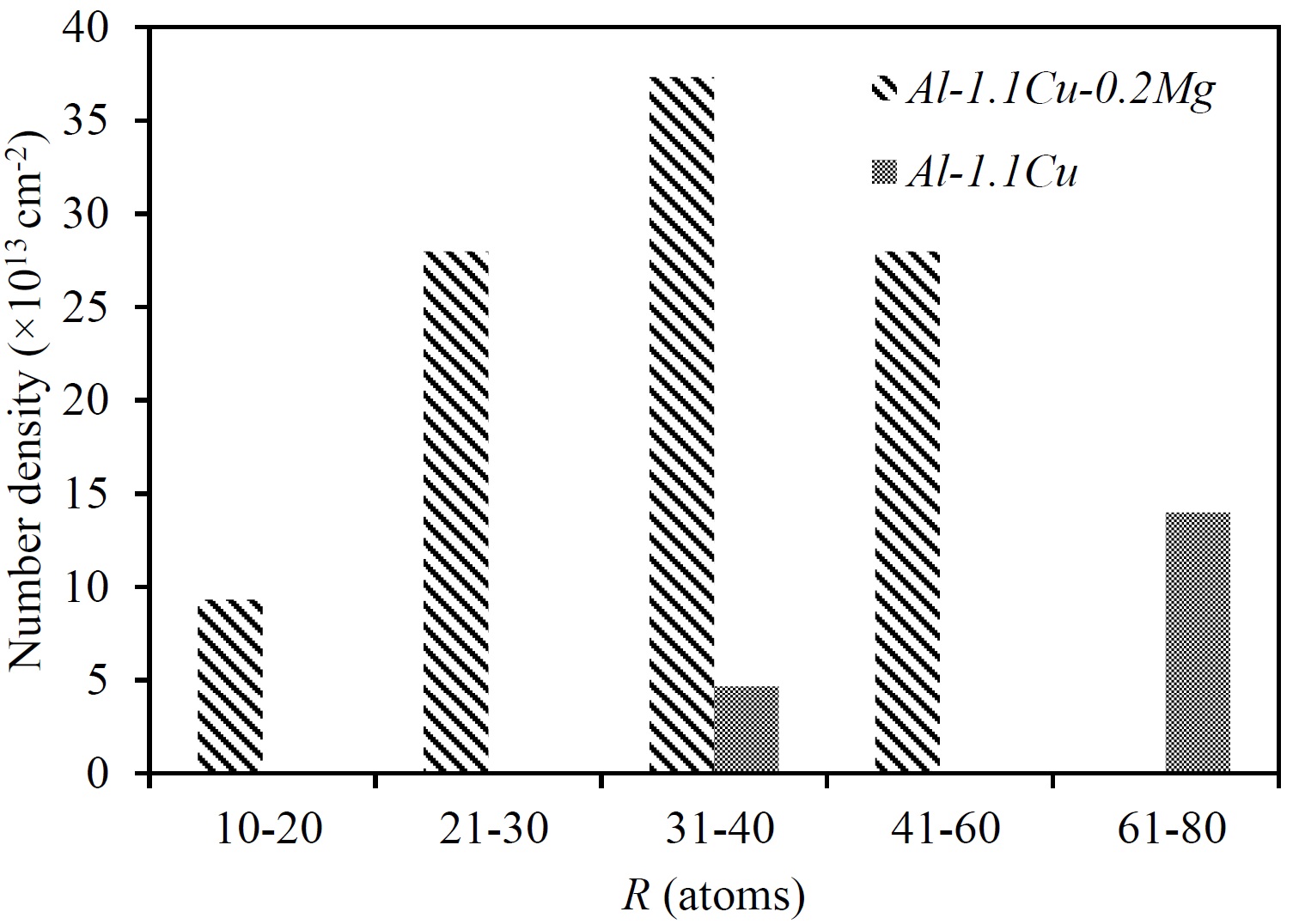}}
\caption{Number density vs. cluster size distribution in solutionized/quenched Al-1.1Cu and Al-1.1Cu-0.2Mg alloys.}
\label{fig:ClusterDensity}
\end{figure}

A typical simulation result is used to study the effect of adding Mg to Al-1.1Cu alloy on the evolution of clusters, as illustrated in Fig.~\ref{fig:Clustering}. The cluster radius $R$, measured in terms of the number of atoms, is defined by radially averaging the radius of the concentration field bound by a threshold of concentration $c_{Cu}^{th}$ defined by  
\begin{align}
c_{Cu}^{th}=c_{Cu}^b+\frac{\sum^N {c_{Cu}-c_{Cu}^b}}{N},
\label{cTh}
\end{align} 
which neglects the small concentrations of Mg. Here, $N$ is the number of mesh points within the selected domain containing the cluster, and superscript '$b$' denotes bulk properties. $c_{Cu}^b$ defines the far field concentration of Cu within the above selected domain. The number density of clusters is estimated by normalising the number of clusters of each size range within the simulation domain with respect to the unit cell area of a 2D square crystal structure with the lattice parameter of $2 r_{fcc}^{Al}=2.86\times10^{-8}cm$. The number density of clusters and their size distribution is plotted in Fig.~\ref{fig:ClusterDensity}, at $t=30,000$ time steps. The data shows that Mg alloying in Al-Cu alloy promotes clustering through an increase in the number density and a reduction in the average size of the clusters, as also noted from Fig.~\ref{fig:Clustering}. 

These simulation results are consistent with a number of experimental observations made on the quenched/naturally-aged Al-Cu-Mg alloys using a combination of TEM~\cite{marceau10-2}, PAS~\cite{marceau10-2,somoza02} and 3D atom probe technique (APT)~\cite{marceau10-2,ringer97}. With APT technique, Ringer {\it et al.}~\cite{ringer97} and Marceau {\it et al.}~\cite{marceau10-2} found pre-precipitate Cu-Mg co-clusters of $\approx$3-20 atoms are distributed within the solid solution matrix at an early-stage ageing of the solutionized and quenched Al-1.1Cu-(0.2-1.7)Mg alloys. In particular, Marceau {\it et al.}~\cite{marceau10-2} noted a marked increase in the number density of clusters of various sizes by raising the Mg-content in these alloys. Notably, the increase in number density was more pronounced for smaller cluster sizes, leading to a smaller average cluster size for alloys with higher Mg-content. Furthermore, using the PAS technique for these alloys in the quenched-state, a significant increase in the positron lifetime was recorded at higher Mg-contents, indicating that Mg may stabilise the free volume in the matrix (i.e., increasing the number density of quenched-in defects), possibly by co-clustering with Cu~\cite{somoza02,marceau10-2}. Somoza {\it et al.}~\cite{somoza02} also observed that in Al-1.74Cu-0.35Mg alloy, immediately after quenching, Cu-content at the positron annihilation site was higher than that in Al-1.74Cu alloy. They also pointed out that Cu-Mg complexes acted as embryos for further aggregation of Cu resulting in higher kinetics of clustering during ageing of the solutionaizd/quenched Al-1.74Cu-0.35Mg alloy compared to that in Al-1.74Cu alloy. Rapid solute aggregation during the early stage of ageing of solutionized/quenched Al-Cu-Mg alloys has been suggested through calorimetric measurements by Zahra {\it et al.}~\cite{zahra98}. 

\subsubsection{Evolution of cluster composition}

\begin{figure}[htbp]
\resizebox{3.3in}{!}{\includegraphics{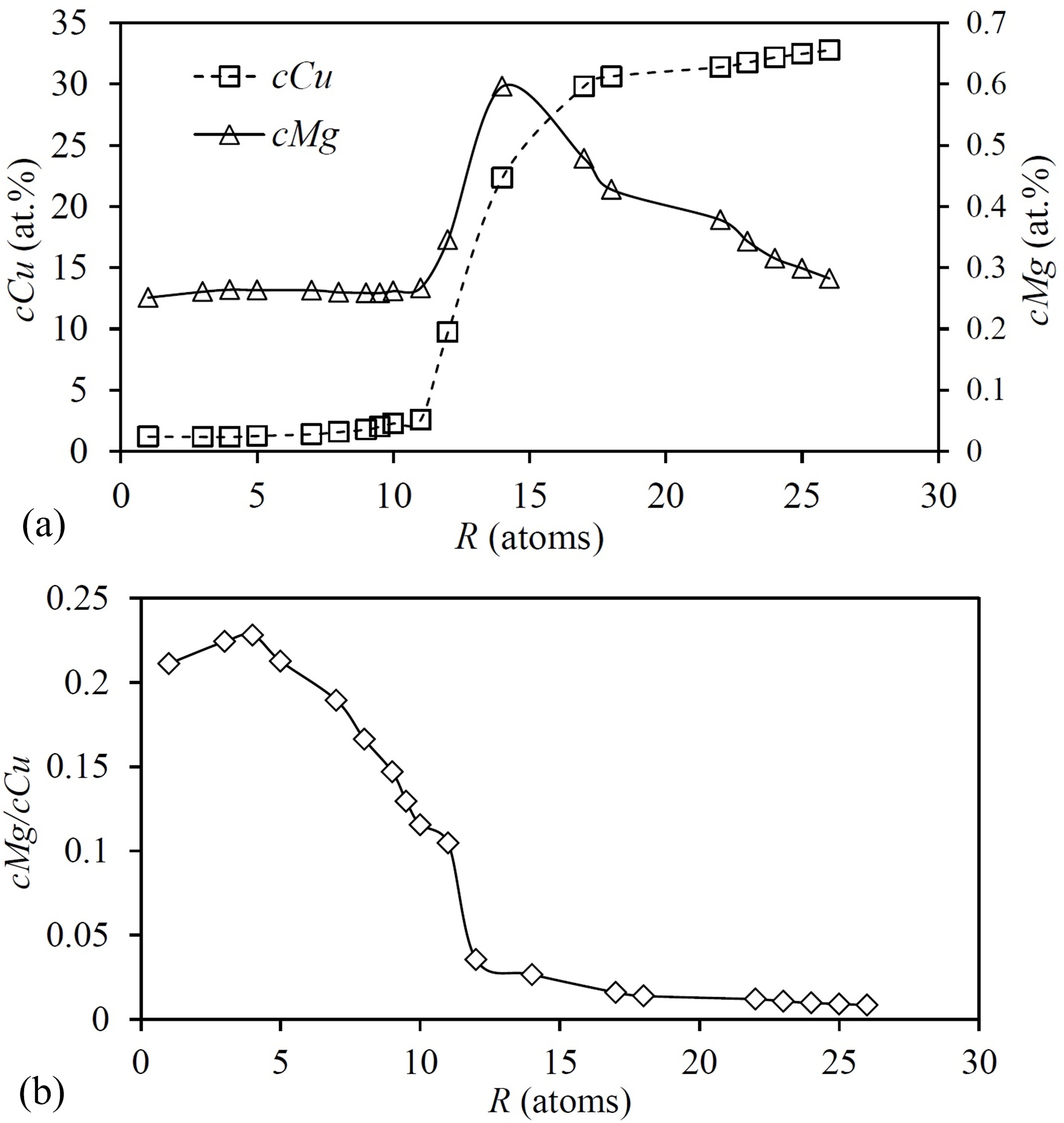}}
\caption{Evolution of (a) composition and (b) Mg/Cu ratio for cluster ``$a'$'' shown in Fig.~\ref{fig:Clustering}(e)-(h). $R=0$ defines the centre of the cluster.}
\label{fig:ClusterComp}
\end{figure}

The compositional evolution of cluster ``$a'$'' (marked on the images shown in Fig.~\ref{fig:Clustering}(d)-(h), as it grows in the solutionized/quenched Al-1.1Cu-0.2Mg alloy, is illustrated in Fig.~\ref{fig:ClusterComp}. The cluster composition is estimated by averaging within a circle of radius $R$, which is determined by a threshold Cu-content (i.e., $c_{Cu}^{th}$ as defined by Eq.~\ref{cTh}). From the data of Fig.~\ref{fig:ClusterComp}, we find that the Cu-content continuously rises towards its equilibrium value in the $\theta$-phase (i.e., $c_{Cu}^{eq}\approxeq 32.8 at.\%$), as specified by the constructed phase diagram (Fig.~\ref{fig:PhaseDiagram}). Meanwhile, Mg-content and Mg/Cu ratio reach their maximum before they continuously drop close to their equilibrium values (i.e., $c_{Mg}^{eq}\approxeq 0.29 at.\%$ and $(\frac{c_{Mg}}{c_{Cu}})_{eq}\approxeq 0.009$) in the $\theta$-phase. Here, the formation process of a Cu-rich cluster with the equilibrium concentration can be divided into two successive steps; First, an increase in both Cu- and Mg-content within the small early clusters, and second, a reduction in the Mg-content while the cluster continues to attract more Cu atoms and grows in size until it forms a stable Cu-rich Cu-Mg cluster. The above evolution of composition during the two-stage clustering is investigated in more depth in the next section through a detailed analysis of the system thermodynamics, which is also supported by relevant data from the existing experiments.

\section{Analysis of Cluster Formation}
\label{En}

Formation of clusters is studied through the analyses of the energy in the system from the earliest small embryos to final stable clusters. First, the energetic mechanisms of compositional evolution is investigated for a small embryo forming around strained regions in the matrix (i.e., around the dislocations displacement fields) until it grows into a stable cluster.   

\subsection{Compositional change from embryo to cluster}

\begin{figure*}[htbp]
\resizebox{6in}{!}{\includegraphics{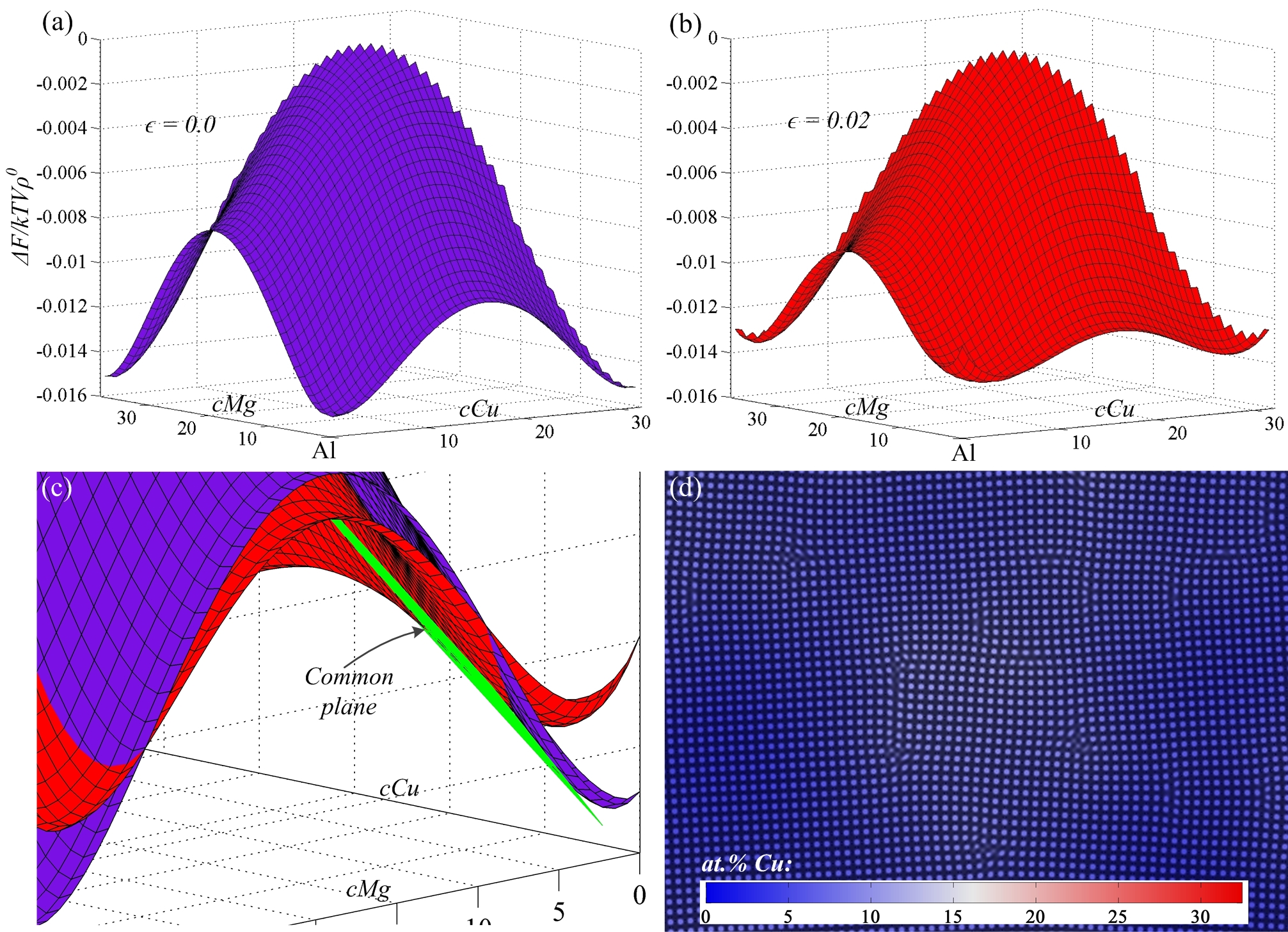}}
\caption{(a) The free energy of the unstrained matrix; (b) The mean-field approximation of the free energy of a uniformly strained matrix at $\epsilon=0.02$; (c) Metastable coexistence between the strain-free matrix (blue surface) and the strained regions (red surface) around the dislocations at a temperature of $\sigma=0.04$; (d) Area around Cluster labelled ``$a'$'' in Fig.~\ref{fig:Clustering}(f) at an early-stage of the transformation, $t=2,000$ time steps.}
\label{fig:StrainedEnergy}
\end{figure*}

The early small clusters, also being called embryos, show an increase in the Mg content and Mg/Cu ratio (as illustrated in Fig.~\ref{fig:ClusterComp}) up to $\approx 14$ atoms in radius. This phenomenon can be qualitatively explained by considering a metastable coexistence between the strain-free bulk matrix and the strained areas near dislocations containing a small cluster (i.e., blue and red energy surfaces, plotted in Fig.~\ref{fig:StrainedEnergy}(a) and (b), respectively). Following the methodology of Fallah {\it et al.}~\cite{fallah12}, the effect of strain on the mean field approximation of the free energy of the system is evaluated in the solid-state, at a given temperature of $\sigma=0.04$, and shown in Fig.~\ref{fig:StrainedEnergy}(b). In this methodology, the magnitude of distortion caused by the displacement fields of dislocations near the cluster is evaluated and mapped onto an effective uniform strain. The strain magnitude is then used for calculating the free energy of the distorted structure shown in Fig.~\ref{fig:StrainedEnergy}(d), which represents an area containing the early-stage development of a cluster (labelled ``$a'$'' in Fig.~\ref{fig:Clustering}(f)). The concentration map of Cu is shown overlaid on the microstructure in Fig.~\ref{fig:StrainedEnergy}(d), revealing the segregation of Cu into the distorted areas near dislocations. The concentration map of Mg also follows the same path qualitatively. The strain value is approximated with the following formula over the triangulated density peaks using the Delaunay Triangulation method:

\begin{align}
\epsilon &= \sum_{i=1}^{N_{tri}} \sum_{j=1}^{3}\bigg(\frac{a_{ij}-a_o}{a_o}\bigg),
\label{Strain}
\end{align} 
where $N_{tri}$ is the number of triangles in the field, $a_o$ is the dimensionless equilibrium lattice parameter (the number of grid points resolving one lattice spacing, i.e., 8), $a_{ij}$ is the length of the $j^{th}$ side of the $i^{th}$ triangle. The magnitude of strain for the particular structure shown on Fig.~\ref{fig:StrainedEnergy}(d) is estimated to be $\epsilon \approx 0.02$. At a given temperature, the strain can be evaluated by calculating the peaks of the correlation kernel $\hat{C}_{2j}^{ii}$ in reciprocal space, at locations $k_j$ defined by a slight deviation from those of the equilibrium density peaks. The following equation defines the amount of strain: 
\begin{align}
\epsilon &= |k-k_j|/k_j,
\label{FourierStrain}
\end{align}
where index $j$ denotes one family of planes in reciprocal space.

For the alloy Al-1.1Cu-0.2Mg, the common tangent plane construction illustrated in Fig.~\ref{fig:StrainedEnergy}(c) reveals a set of compositions for a coexisting unstrained matrix and strained solid in the vicinity of dislocations. Particularly, it denotes higher concentrations of Mg and Cu in the strained areas than those in the bulk unstrained matrix, while the two are in a metastable coexistence. Such phenomenon of segregation of solute atoms into the strained areas has been shown to be driven by a stress relaxation mechanism~\cite{fallah12,leonard05,muralidharan10}. Fallah {\it et al.}~\cite{fallah12} have shown that this mechanism reduces and/or even completely eliminates the energy barrier for formation of stable clusters in binary alloys.

The above phenomenon is responsible for the first rise in the Mg-content illustrated in Fig.~\ref{fig:ClusterComp}(a). It can be argued that initially both Mg and Cu atoms segregate into nearby dislocations to relax their stress fields while also forming a small Cu-Mg co-cluster. The cluster composition estimated from the above metastable  coexistence (i.e., $c_{Cu}\approxeq 6.7at.\%$ and $c_{Mg}\approxeq 0.7at.\%$) is similar to that measured for cluster ``$a'$'' at the peak of Mg-content (see Fig.~\ref{fig:ClusterComp}(a)). The first rise in the Mg-content and thus in the Mg/Cu ratio is consistent with the atom probe data of Marceau {\it et al.}~\cite{marceau10-2} showing an increase in the Mg/Cu ratio by the size of clusters up to $\approx$13 atoms in the naturally aged Al-1.1Cu-(0.2-1.7)Mg alloys. Also, the PAS investigation of clustering in Al-1.74Cu-0.35Mg and Al-1.7Cu-1.3Mg by Somoza {\it et al.}~\cite{somoza00} and Nagai {\it et al.}~\cite{nagai01}, respectively, showed that, right after quenching, a high volume of the bulk crystal defects are associated with solute atoms~\cite{somoza00,nagai01}.

In the second stage of clustering, while a randomly selected small cluster continues to grow among the other clusters, it attracts more Cu atoms causing the Mg-content to decrease (see Fig.~\ref{fig:ClusterComp}(a)). This occurs since the growing cluster releases more stress from the matrix by attracting more Cu atoms and moving towards the composition of the equilibrium $\theta$-phase, which contains less Mg (i.e., $\approxeq 0.29at.\%$, referring to the phase diagram shown in Fig.\ref{fig:PhaseDiagram}(d)) than the highly-strained initial Cu-Mg co-cluster. This phenomenon is in accordance with the experimental evidence of the increase in the local Cu-content at the positron annihilation sites (i.e., vacancy-Cu-Mg complexes) during ageing of the solutionized/quenched Al-1.74Cu-0.35Mg alloy~\cite{somoza02,somoza00}. Moreover, the PAS investigation of clustering during early-stage ageing of Al-1.74Cu-0.35Mg~\cite{somoza00} and Al-1.7Cu-1.3Mg~\cite{nagai01} alloys showed that the bulk crystal defects mingle more effectively with Cu-Mg~\cite{somoza00,nagai01} aggregates, which are considered as more efficient positron traps than the single solute atoms. 

The above results and analysis suggest that although the crystal defects are strongly attached to the solute atoms (i.e., both Cu and Mg) immediately after quenching, they associate more effectively with Cu atoms than Mg during the initial growth of clusters upon ageing~\cite{somoza00}. More specifically, Nagai {\it et al.}~\cite{nagai01} report that defects are more effectively bound to Mg atoms rather than Cu atoms in the as-quenched state. However, during the early stages of the subsequent ageing, they observe clustering of Cu-rich Cu-Mg complexes along the dislocations.   

\subsection{Work of formation}

In order to investigate the effect of trace addition of Mg on the clustering behaviour of a quenched$/$aged dilute Al-Cu-Mg alloy, we first analyse the work of  formation of a long-lived cluster. The work of formation for clustering is defined as
\begin{align}
W_h &= 2\pi R\gamma + \pi R^2 (-\Delta f + \Delta G_s)
\label{Hom.Nucl.En}
\end{align}
where $R$ is the cluster radius in terms of number of lattice spacings, $\Delta f$ is the bulk driving force for nucleation of a cluster at a given concentration, $\Delta G_s$ represents the strain energy for a coherent nucleus and $\gamma$ is the interfacial free energy per unit length of the interface. We consider 2D in this work but we expect the same mechanisms reported below to hold in 3D. Particularly, our preliminary simulations in 3D have yilded qualitatively similar results, which will be presented in a sequel study. The above quantities are estimated below for clustering in Al-Cu-Mg alloys lying within the two-phase region of $(Al)$-$\theta$. The evolving clusters are then Cu-rich, close in concentration and lattice parameter to those of the equilibrium $\theta$ phase, as shown on the calculated phase diagram in Fig.~\ref{fig:PhaseDiagram}(d).  

\subsubsection{Driving force for the formation of clusters}

\begin{figure}[htbp]
\resizebox{3.3in}{!}{\includegraphics{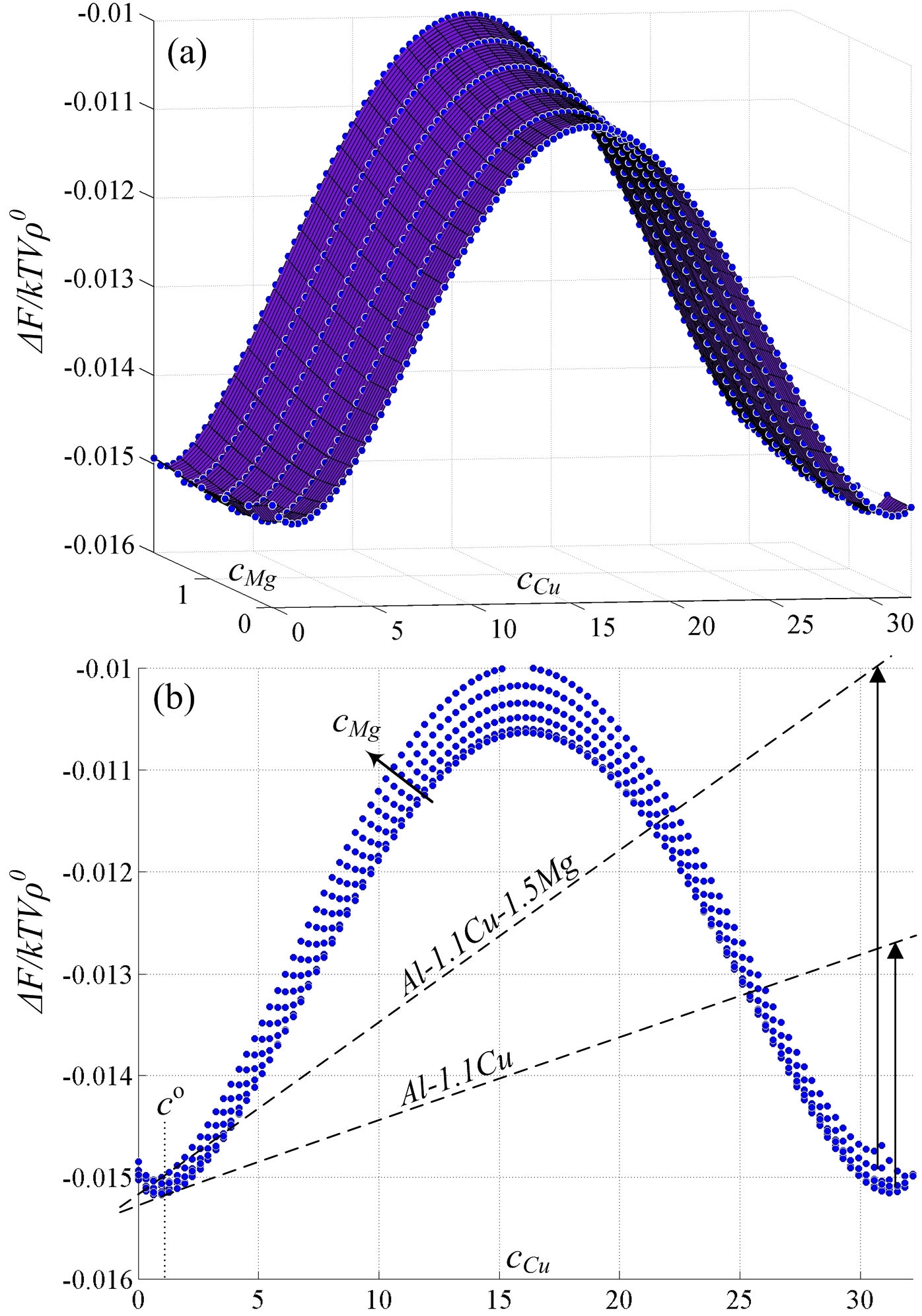}}
\caption{(a) $(Al)$-$\theta$-$\beta$ solid-phase free energy diagram in dilute Mg region; (b) Schematic illustration of calculation of the driving force for clustering of a Cu-rich Cu-Mg co-cluster at two different levels of Mg content.}
\label{fig:DrivingForce}
\end{figure}

The bulk driving force for the formation of a Cu-rich Cu-Mg co-cluster with the equilibrium concentration is defined as 
\begin{align}
-\Delta f = f^b - \mu_{c_{Cu}}^b|_{c_{Cu}^b}(c_{Cu}^b-c_{Cu}^{cl})\nline - \mu_{c_{Mg}}^b|_{c_{Mg}^b}(c_{Mg}^b-c_{Mg}^{cl})-f^{cl},
\label{DrivingForce}
\end{align}
where superscripts `$b$' and `$cl$' denote the bulk matrix and cluster ``phase'' quantities, respectively. Fig.~\ref{fig:DrivingForce}(a) shows a low-Mg section of $(Al)$-$\theta$-$\beta$ solid-phase free energy diagram. Using this free energy diagram, we compute the approximate driving force for clustering of a Cu-rich Cu-Mg co-cluster at two different levels of Mg-content (see Fig.~\ref{fig:DrivingForce}(b)). The illustration of these calculations on Fig.~\ref{fig:DrivingForce}(b) excludes the effect of $\mu_{c_{Mg}}^b$ in Eq.~\ref{DrivingForce} on the driving force, assuming the overall difference in the Mg content between the matrix and final Cu-rich cluster is negligible. It is seen that Mg alloying increases the driving force for clustering of the Cu-rich clusters (see also Fig.~\ref{fig:StrainEnergy}(c)). In other words, addition of a small amount of Mg into the Al-1.1Cu alloy decreases the solubility of Cu (i.e., from $\approx 0.8$ to $\approx 0.6$ $at.\%$Cu by adding 0.2$at.\%$Mg, obtained from the common tangent construction in Fig.~\ref{fig:DrivingForce}(b)), thus raising its supersaturation in the quenched state.  

\subsubsection{Strain energy created by clusters}

The strain energy for a coherent nucleus is evaluated by~\cite{hoyt}:

\begin{align}
\Delta G_s = 2 G_A \delta^2 \frac{K_B}{K_B+G_A},
\label{StrainEnergy}
\end{align}
where 
\begin{align}
\delta = (a_{Cu}-a_{Al})(c_{Cu}-c_{Cu}^b) + (a_{Mg}-a_{Al})(c_{Mg}-c_{Mg}^b)
\label{misfit}
\end{align}
is the misfit strain and $G_A$ and $K_B$ are 2D shear and bulk moduli, respectively, calculated from PFC 2D mode approximation~\cite{greenwood11}. Following Greenwood {\it et al.}~\cite{greenwood11}, in the limit of small deformations~\cite{elder10}, the free energy of different strained states in a 2D crystal of square symmetry is evaluated through substituting their respective coordinate transformations into a two mode approximation of the density field and integrating over the bounds of the strained unit cell. The elastic constants $C_{11}$, $C_{12}$ and $C_{44}$ ($C_{12}=C_{44}=C_{11}/3$) are then extracted through fitting the resultant free energy to parabolic expansions in displacement fields. In two dimensions, the shear and bulk moduli can be simply defined as $G_A=C_{44}$ and $K_B=\frac {C_{11}+C_{12}}{2}$ (radially averaged for square symmetry~\cite{greenwood11}), respectively~\cite{elder04}.        

\begin{figure*}[htbp]
\resizebox{5.5in}{!}{\includegraphics{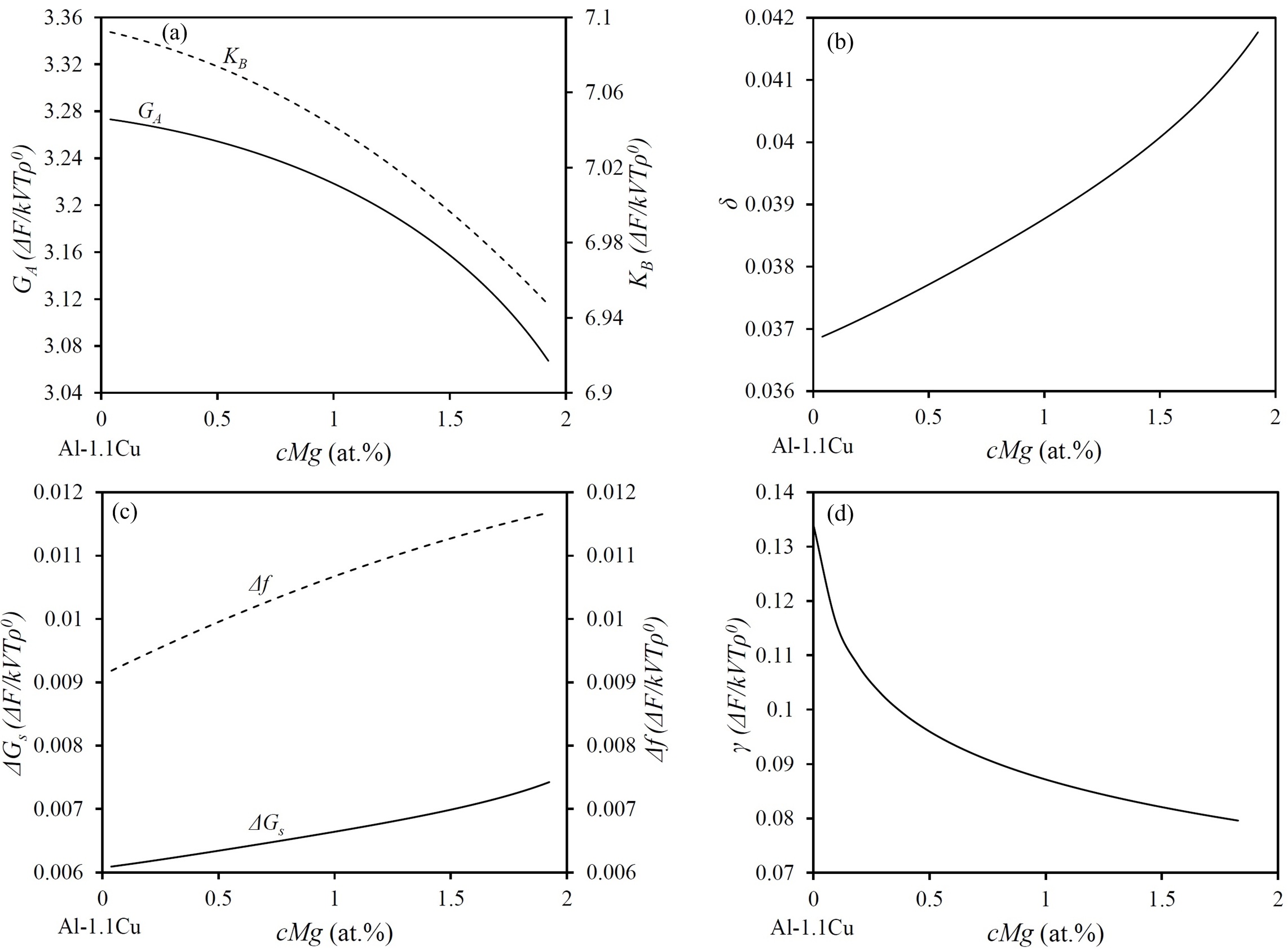}}
\caption{The effect of Mg on (a) the shear and bulk moduli, $G_A$ and $K_B$, respectively, (b) the misfit strain $\delta$, (c) the bulk strain energy, $\Delta G_s$, and the driving force of formation, $\Delta f$, and (d) the surface energy associated with Cu-rich Cu-Mg co-clusters in an Al-rich matrix of Al-1.1Cu-$x$Mg alloys.}
\label{fig:StrainEnergy}
\end{figure*}

We calculate the bulk modulus for the Cu-rich cluster of the equilibrium concentration and the shear modulus for the Al-rich matrix for varying Mg-content, according to the approach presented in Ref.~\cite{hoyt}. Figures~\ref{fig:StrainEnergy}(a) and (b) show the effect of Mg alloying on the elastic moduli of Al-1.1Cu-$x$Mg alloys and the misfit strain, $\delta$, respectively, created by clustering of a Cu-rich cluster of the equilibrium concentration. For higher Mg contents, the combination of a reduction in the bulk and shear moduli of the matrix and an increase in the misfit strain created by the formation of a Cu-rich cluster results in higher strain energy values (as illustrated in Fig.~\ref{fig:StrainEnergy}(c)).

\subsubsection{Surface energy of clusters}

Assuming a low dislocation density in the system, the interfacial free energy is taken to be solely chemical, neglecting the structural contributions~\cite{Turnbull}. Following Cahn and Hilliard~\cite{cahn58}, the interfacial energy between a Cu-Mg co-cluster and the Al-rich matrix is evaluated by the following analytical form which uses the composition-dependent mean-field free energy, $f(c_{Cu},c_{Mg})$,:   
\begin{align}
\gamma &= 2\int\limits_{c_{Cu}^b}^{c_{Cu}^{cl}}{\bigg[\alpha_{Cu} (f-f^b)\bigg]^\frac{1}{2}\bigg\{1+(\frac{\alpha_{Mg}}{\alpha_{Cu}})(\dxdy{c_{Mg}}{c_{Cu}})^2\bigg\}^\frac{1}{2}}dc_{Cu},
\label{Surf.E}
\end{align}
where $\alpha_{Cu}$ and $\alpha_{Mg}$ are gradient energy coefficients for Cu and Mg, respectively, both set to 1 in this study. The term $\dxdy{c_{Mg}}{c_{Cu}}$ is estimated by $-\frac{\frac{\partial f}{\partial c_{Cu}}}{\frac{\partial f}{\partial c_{Mg}}}$ and the variation of $c_{Mg}$ with respect to $c_{Cu}$ is approximated by a linear interpolation between the bulk and the equilibrium cluster compositions (i.e., $c_{Mg}=c_{Mg}^b+\frac{c_{Mg}^{cl}-c_{Mg}^b}{c_{Cu}^{cl}-c_{Cu}^b}(c_{Cu}-c_{Cu}^b)$). As illustrated in Fig.~\ref{fig:StrainEnergy}(d), increasing the amount of Mg in Al-1.1Cu-$x$Mg alloys decreases the surface energy of the Cu-rich clusters. 

\begin{figure*}[htbp]
\resizebox{6.5in}{!}{\includegraphics{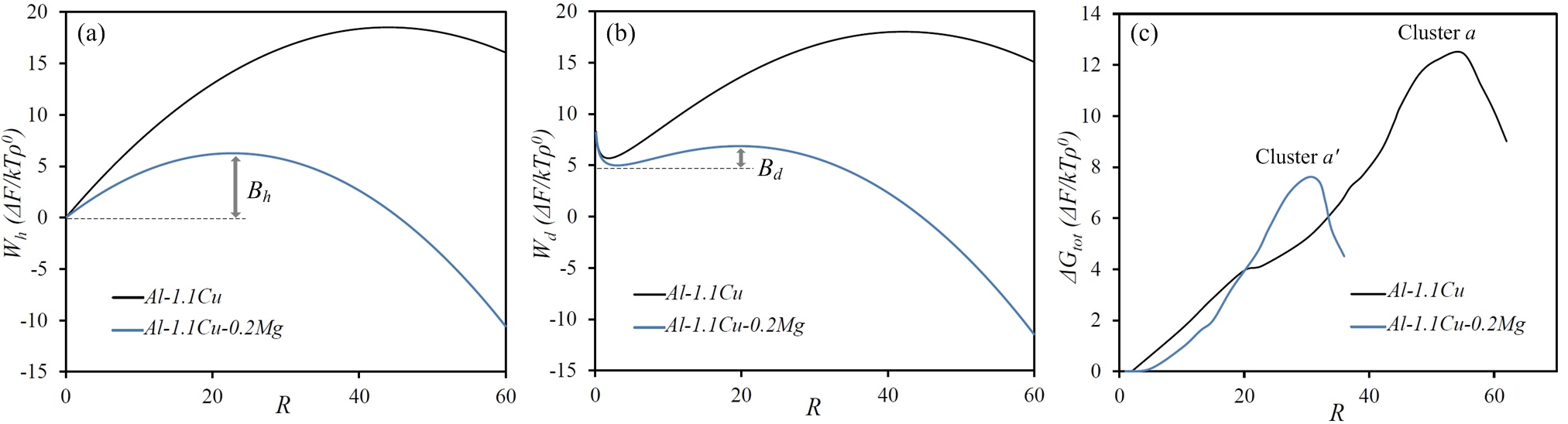}}
\caption{Work of formation for (a) coherent/homogeneous (evaluating Eq.~\ref{Dis.Nucl.En}) and (b) dislocation-assisted (evaluating Eq.~\ref{Hom.Nucl.En} for one dislocation, i.e., $\Sigma b_i^2=1$) clustering of Cu-rich Cu-Mg co-clusters in an Al-rich matrix of Al-1.1Cu-$x$Mg alloys; (c) The variation of numerically evaluated total energy, $\Delta G_{tot}$, due to the formation of clusters ``$a$" and ``$a'$'', within the boxes marked on the images shown in Fig.~\ref{fig:Clustering}(a)-(d) and (e)-(h), respectively.}
\label{fig:NucleationEnergy}
\end{figure*}

Fig.~\ref{fig:NucleationEnergy}(a) depicts the evaluation of the work of formation in Eq.~\ref{Hom.Nucl.En}, which combines the effects of the computed values of the surface and strain energy and the driving force of clustering. As can be inferred from this figure, the addition of Mg to the Al-1.1Cu alloy lowers the energy barrier height and the critical size of nuclei for the formation of a stable Cu-rich Cu-Mg cluster.    

\subsubsection{Effect of dislocations}

The effect of dislocations on the formation of clusters can be described augmenting the work of formation by the following form~\cite{fallah12}
\begin{align}
W_d &= W_h - \Delta G_{sr} + \Delta G_d \nline&= 2\pi R\gamma + \pi R^2 (-\Delta f + \Delta G_s) - \Delta G_{sr} + \Delta G_d
\label{Dis.Nucl.En}
\end{align}
$\Delta G_{sr}$ is the stress relaxation term due to segregation of solute into dislocations~\cite{cahn57}, described here by
\begin{align}
\Delta G_{sr} = \eta^2\chi_d E A \ln(R),
\label{StressRelaxE}
\end{align}
where $A= \frac{G_A \Sigma b_i^2}{4 \pi (1-\nu)}$, $\nu = \frac{E}{2 G_A}-1$, $E=\frac{4K_B G_A}{K_B+G_A}$ is the 2D Young's modulus, $\eta$ is the linear expansion coefficient with respect to concentration, $\chi_d$ represents the change in the diffusion potentials due to concentration, and we introduce $\Sigma b_i^2$ to represent a weighted average of the magnitude of Burger's vectors around the dislocations accompanying the cluster and $a$ is the lattice parameter. The prefactor $\eta^2\chi_d E A$ accounts for the reduction in strain energy due to solute segregation around a dislocation~\cite{larche85}. Following the work of Cahn and Larche~\cite{larche85}, which approximates the analysis of King {\it et al.}~\cite{king91} and works well for dilute solutions, we obtain the following approximation for the term $\eta^2\chi_d$ in an isotropic ternary system  
\begin{align}
\eta^2 \chi_d = \frac{\eta_{Mg} \eta_{Mg} \frac{\partial^2 f}{\partial c_{Cu}^2} + \eta_{Cu}\eta_{Cu} \frac{\partial^2 f}{\partial c_{Mg}^2} +2 \eta_{Cu} \eta_{Mg} \frac{\partial^2 f}{\partial c_{Cu} \partial c_{Mg}}}{\frac{\partial^2 f}{\partial c_{Cu}^2}\frac{\partial^2 f}{\partial c_{Mg}^2}-\left(\frac{\partial^2 f}{\partial c_{Cu} \partial c_{Mg} }\right)^2},
\label{etachi}
\end{align}
where $\eta_{Cu}=-\frac{1}{a}\frac{\partial a}{\partial c_{Cu}}$ and $\eta_{Mg}=-\frac{1}{a}\frac{\partial a}{\partial c_{Mg}}$. The estimated values of the term $\eta^2\chi_d$ show an increase from 0.0547 to 0.0823 for the bulk matrix of Al-1.1Cu and Al-1.1Cu-0.2Mg alloys, respectively. This implies that adding Mg to the Al-1.1Cu alloy enhances the stress relaxation within the distorted matrix through segregation of solutes (both Cu and Mg) into areas near the dislocations.  

The term $\Delta G_d$ in Eq.(\ref{Dis.Nucl.En}) accounts for the increase in the total energy of the system due to presence of dislocations. Its form is approximated by  
\begin{align}
\Delta G_d = \zeta A,
\label{Disl.E}
\end{align}
where $\zeta$ is a prefactor of the order 10, representing the average amount of energy per dislocation core~\cite{Hull}. 

Eq.(\ref{Dis.Nucl.En}) for the work of formation for dislocation-assisted clustering in Al-1.1Cu and Al-1.1Cu-0.2Mg alloys is plotted in Fig.~\ref{fig:NucleationEnergy}(b). This figure shows that for the case of a strained area around one dislocation (i.e., $\Sigma b_i^2=1$) the energy barrier height for clustering of Cu-rich Cu-Mg co-clusters of any Mg-content is lower than that of the homogeneous clustering (as depicted in Fig.~\ref{fig:NucleationEnergy}(a) and (b) with barrier heights labelled $B_h$ and $B_d$ for homogeneous and dislocation-mediated clustering, respectively. 

The total work of formation, $\Delta G_{tot}$, was also computed numerically by measuring the change in the grand potential within a box engulfing clusters ``$a$" and ``$a'$'' during their formation and growth in the bulk matrix, i.e., 
\begin{align}
&\Delta G_{tot}=\int_V\omega-\int_V\omega^b=\nline&\int_V{[f-\mu_{c_{Cu}}.c_{Cu}-\mu_{c_{Mg}}.c_{Mg}-\mu_n.n]}\nline&-\int_V {[f^b-\mu_{c_{Cu}}^b.c_{Cu}^b-\mu_{c_{Mg}}^b.c_{Mg}^b-\mu_n^b.n^b]}.
\label{TotalE}
\end{align}
Here, $\mu_{c_{Cu}}=\frac{\delta \check{{\cal F}}}{\delta {c_{Cu}}}$, $\mu_{c_{Mg}}=\frac{\delta \check{{\cal F}}}{\delta {c_{Mg}}}$ and $\mu_n=\frac{\delta \check{{\cal F}}}{\delta n}$ are diffusion potentials of concentration and density fields, respectively, and $V$ is the total volume. The total work of formation, $\Delta G_{tot}$, has contributions from the surface energy and driving force for formation of clusters (i.e., $\Delta G_{tot}=\Delta G_{\gamma}-\Delta G_v$), both including also the elastic effects. As can be seen in Fig.~\ref{fig:NucleationEnergy}(c), the total work of formation increases with the growth of clusters to a maximum value and then decreases. The peak in the total work of formation can be explained in such a way that the measuring boxes (marked on the images shown in Fig.~\ref{fig:Clustering}) contain only one growing cluster, i.e., cluster ``$a$" or ``$a'$''. Therefore, the calculated change in the grand potential accounts for structural and compositional changes during the formation and growth of the targeted cluster. It should be noted that  while the growth of one cluster may raise the local free energy, other parts of the system can undergo a process of annihilation and/or shrinkage of sub-critical clusters, thus  resulting in a decrease of the total free energy of the system.

According to our simulation data, in the presence of multiple dislocations, typical clusters ``$a$" or ``$a'$'' continuously grow until they become stable. This suggests that at each sub-critical cluster size (i.e., below $\approx 30$ and $\approx 50$ atoms in radius, equivalent to the radius of clusters ``$a$" and ``$a'$'', respectively, at the peak of energy shown in Fig.~\ref{fig:NucleationEnergy}(c)), the system is sitting at a local energy minimum. Consistent with the previously proposed mechanism~\cite{fallah12}, locally straining a sub-critical cluster due to local accumulation of the magnitudes of a collection of dislocations Burger's vectors creates a sufficiently large $\Sigma b_i^2$ (as illustrated in Fig.~\ref{fig:ClusteringStructure}(b)), which eventually leads to complete elimination of the energy barrier. It is thus thermodynamically favorable for the sub-critical clusters to accumulate solute atoms from the matrix and grow in size. However, since the accumulation of a sufficient number of dislocations into the neighbourhood of a cluster is a statistical occurrence, not all sub-critical clusters will be long-lived, and  some of them may even reverse their growth and eventually disappear (if not grow beyond the critical size).

Past the critical size of the stable cluster, even the local free energy associated with the growing cluster starts to decrease due to the dominant role of the driving force over the interfacial energy. The observed peak in the numerically evaluated total work of formation, $\Delta G_{tot}$, is consistent with the estimated energy barriers obtained through analysis of the work of formation for homogeneous and dislocation-assisted clustering (see plots in Fig.~\ref{fig:NucleationEnergy}(a) and (b)). One can easily conclude that addition of Mg to the Al-1.1Cu alloy reduces energy barrier and the critical size of the stable cluster thus leading to a higher clustering rate and a finer distribution of clusters.    

\section{Conclusion}
In summary, we utilized a ternary extension of the alloy phase field crystal model of ref.~\cite{greenwood11-2} to simulate and analyze the atomistic mechanisms governing the early-stage clustering phenomenon in ternary alloys. Our previous energy analysis~\cite{fallah12} of dislocation-mediated clustering in the binary Al-Cu system was extended to include the effects from adding a ternary element. Particularly, consistent with the existing experiments, we showed that Mg alloying in Al-Cu-Mg system refines the final microstructure. 

The detailed analysis of the system energetics for different levels of Mg-content in quenched/aged Al-Cu-Mg alloys revealed that the addition of Mg increases the effective driving force for nucleation (i.e., $\Delta f - \Delta G_s$), decreases the surface energy, $\gamma$, and enhances the dislocation stress relaxation, $\Delta G_{sr}$, associated with the Cu-rich Cu-Mg co-clusters. This in turn ensured a higher rate of nucleation thus leading to a finer distribution of clusters, a phenomenon which was also observed experimentally~\cite{marceau10-2,somoza02}. Furthermore, we showed that the simulation results for the compositional evolution of the early clusters are in accordance with the previously obtained chemical composition data through PAS~\cite{marceau10-2,somoza02,nagai01} and 3D APT~\cite{marceau10-2} analyses. Particularly, through free energy analysis of the areas engulfing the typical stable clusters, we showed that it is thermodynamically favourable for the small (sub-critical) early clusters to contain more Mg than the larger (over-critical) Cu-rich clusters. Such phenomenon led us to define a two-stage clustering in Al-Cu-Mg system, comprising an initial enrichment of the cluster in both Cu and Mg followed by attraction of only Cu atoms leading to a reduction in the Mg-content, while also
moving towards  equilibrium concentrations in both components. 

To our knowledge, this is the first ternary phase field crystal (PFC) model that elucidates the role of ternary elements on the phenomenon of clustering. Through a self-consistent coupling of the diffusive and elasto-plastic effects, our simulations have shed light on the poorly understood atomistic mechanisms through which the chemical composition impacts on the early-stage clustering phenomena in ternary alloys, which are governed by crystal defects and their elastic interactions. We expect our results to hold qualitatively in 3D. Particularly, our preliminary 3D simulations of clustering in dilute Al-Cu alloys reveal dislocation-mediated formation and growth of early clusters, the results of which will be presented in a sequel study. The energy-based methodology presented in this study will be extended to 3D in the future to include more complex crystal structures and/or chemical compositions. Finally, while we have investigated the clustering behaviour in the presence of dislocations, we expect that the salient features of the mechanisms proposed in this work will hold for vacancy-assisted clustering and ageing in multi-component alloys.  

\begin{acknowledgements}
We acknowledge the financial support received from National Science and Engineering Research Council of Canada
(NSERC), Ontario Ministry of Research and Innovation (Early Researcher Award Program), the University of Waterloo and the Clumeq High Performance Centre.
\end{acknowledgements}

\bibliography{references}

\end{document}